\documentclass[twocolumn,aip,cha,showpacs,superscriptaddress,showkeys]{revtex4}

\usepackage{graphicx}
\usepackage{epsfig}
\usepackage{times}
\usepackage{xcolor}   
\usepackage{amsfonts}
\usepackage{amsmath}
\usepackage{amssymb}
\usepackage{amsthm}
\usepackage{hyperref} 
\usepackage{wasysym}
\usepackage{bm} 

\begin{document}

\title{Reliability of UPO based control strategies in biological systems}

\author{Nagender Mishra}
\affiliation{Department of Physics \& Astrophysics, University of Delhi, Delhi 110007, India}
\author{Maria Hasse}
\affiliation{Institut f$\ddot{u}$r H$\ddot{o}$chstleistungsrechnen, Universit$\ddot{a}$t Stuttgart, D-70569, Stuttgart, Germany}
\author{B. Biswal}
\affiliation{Cluster Innovation Center, University of Delhi, Delhi 110007, India}
\affiliation{Sri Venkateswara College, University of Delhi, Delhi 110021, India}
\author{Harinder P. Singh}
\affiliation{Department of Physics \& Astrophysics, University of Delhi, Delhi 110007, India}

\begin{abstract}
Presence of recurrent and statistically significant unstable periodic orbits (UPOs) in time series 
obtained from biological systems  are now routinely used as evidence for low dimensional chaos 
. Extracting accurate dynamical information from the detected UPO 
trajectories are vital for successful control strategies that either aim to stabilize the system 
near the fixed point or steer the system away from the periodic orbits. A hybrid UPO 
detection method from return maps that combines 
topological recurrence criterion, matrix fit algorithm and stringent criterion for fixed 
point location gives accurate and statistically significant UPOs even in the presence of 
significant noise.  Geometry of the return map, frequency of UPOs visiting the same 
trajectory, length of the data set, strength of the noise and degree of nonstationarity affect 
the efficacy of the proposed method. Results suggest that establishing determinism
from unambiguous UPO detection is often possible in short data sets with significant noise, but 
derived dynamical properties are rarely accurate and adequate for controlling the dynamics 
around these UPOs. A repeat chaos control experiment on epileptic hippocampal 
slices through more stringent control strategy and adaptive UPO tracking is reinterpreted 
in this context through simulation of similar control experiments on an analogous but stochastic 
computer model of epileptic brain slices. Reproduction of equivalent results  suggests that far 
more stringent criteria are needed for linking apparent success of control in such experiments 
with possible determinism in the underlying dynamics.
\end{abstract}

\pacs{05.45.-a, 05.45.Gg, 05.45.Tp, 87.18.Bb,87.18.Sn, 87.19.La }
\keywords{Hippocampus, Epilepsy, Unstable periodic orbit}

\maketitle

\textbf{Unstable periodic orbits (UPOs) analysis has now become an 
established tool for detecting determinism in biological systems.
If the system is truly deterministic, accurate UPO characteristics can 
be further used for application of control or anti control strategies
 with potential medical application. However, accurate biological time
series analysis for UPOs critically depend on the length of data set, 
strength of noise, type of geometry, nonstationarity etc. In this work we 
show that it is possible to establish determinism through UPO analysis in 
presence of all these factors. However, the inaccuracy in the measured UPO 
parameters and non-stationarity nature of biological systems mean only adaptive control strategies 
are useful for successful chaos control in such systems. At the same time, comparable success 
of adaptive control strategies on a stochastic neural network model of epileptic slices demand more 
stringent criteria for linking adaptive control success with determinism.}

\section{\label{intro}Introduction}
Chaos control---a method for maintaining periodic dynamics in chaotic systems---has received 
considerable attention following successful experimental demonstration in physical, chemical and
biological systems \cite{Ditto-Showalter1997}.  Applications of
chaos control techniques \cite{Geltrudeetal2012,Santos-Graves2010} in neural dynamics, mainly due
to its possible medical applications in the treatment of epilepsy
\cite{Schiffetal1994}, have also been explored for some time.  A number of
anti-control of chaos methods have also been developed for this purpose
\cite{Schiffetal1994,Inetal1995a,Inetal1998,Sinha-Ditto2001,Guptaetal2006,Iasemidis-Sackellates1996}.  Establishing the occurrence of
deterministic chaos in epilepsy is of fundamental importance. Along
with the possibility of understanding the underlying neuronal dynamics
\cite{Raiesdana-Goplayegani2013,Jiaetal2012,Braunetal2011,Channelletal2009} it also opens the possibility of short term prediction and
control that has potential medical applications.

The first experimental demonstration of chaos control in epilepsy
on spontaneously bursting epileptic hippocampal slices by Schiff {\it{et al.}}  \cite{Schiffetal1994}
 has triggered most of the research activity in this direction.  In this experiment, occurrence of
recurrent unstable periodic orbits (UPOs) like trajectories and
apparent success of chaos control based on these trajectories were
interpreted as evidence for chaotic dynamics.  This was immediately
challenged with application of the same method with comparable success
on a noise driven non-chaotic FitzHugh-Nagumo neuronal model
\cite{Christini-Collins1995}. To resolve this, new surrogate time series analysis
techniques were developed \cite{Soetal1997,Soetal1998,Feietal2012} for both searching the UPOs
in a time series and for assigning  statistical significance to them. 
Application of this method on the inter burst interval (IBI) data sets
from brain slices \cite{Guptaetal2006} and on human epileptic electroencephalogram (EEG)
for statistically significant UPOs reported clear evidence of
determinism \cite{Quyenetal1997}.  However, a rigorous simulation of the chaos control experiment on a stochastic neural
network model for epileptic brain slices could also successfully
reproduce most of the observations reported in the experimental study by Schiff {\it{et al.}} \cite{Biswal-Dasgupta2002a}. 
The surrogate methods for statistically significant UPOs, when applied
to stochastic burst intervals obtained from this computer model, also
produced results \cite{Biswal-Dasgupta2002b} comparable to those reported earlier
\cite{Soetal1998, Quyenetal1997}.  Due to these conflicting results, the issue of
unambiguously establishing deterministic chaos in epilepsy based on 
statistically significant UPOs remained inconclusive.

The chaos control experiment on epileptic hippocampal slices has been
subsequently repeated by Slutzky {\it{et al.}}
\cite{Slutzkyetal2003,Slutzkyetal2002,Slutzkyetal2001} adopting more stringent UPOs detection
method \cite{Iasemidis-Sackellates1996} along with a more robust adaptive chaos control
technique \cite{Christini-Kaplan2000}.  Successful chaos control, as well as
statistically significant UPO analysis of the time series from this
experiment claim to re-establish the original conclusions of low
dimensional deterministic chaos in epileptic brain slices. Various other methods
have been developed to distinguish between deterministic chaos and 
noise \cite{Gaoetal2012a,Gaoetal2012b,Fuh2009,Gao-Jin2012,Yangetal2009} and applied to EEG data to detect seizure \cite{Gaoetal2012a}. 
These positive and promising results, however, have not lead to any real
application so far.  This is because strong skepticism continues
to persist on the two main issues:

\newenvironment{MYitemize}{%
\renewcommand{\labelitemi}{\labelitemii}%
\begin{itemize}}{\end{itemize}}

\begin{MYitemize}
\item whether the apparent UPOs in epileptic EEG truly originate from an
  underlying deterministic dynamics that can be reconstructed or used, and

\item whether the apparent success of any  control strategy using the reconstructed dynamics
      can be conclusively interpreted as due to the deterministic nature of the underlying dynamics.
\end{MYitemize}

The difficulties in providing a clear and convincing answer to these
two issues in the context of epilepsy are numerous.The underlying dynamics of an epileptic brain slice is unknown and
high-dimensional, even the smallest of brain slices contain tens of
thousands of neurons, whereas the applied tools of non-linear dynamics
and deterministic chaos were only proven on low dimensional known
chaotic systems.

The chaos control experiments are based on UPOs extracted from short
inter burst intervals (IBIs).  The apparent geometry visible in the
return maps of truly chaotic systems is absent in the return maps of
the epileptic IBIs in all the reported works.  Interspike time intervals from rat facial
cold receptor for which statistical significant UPOs are established show return
map geometry consistent with UPO trajectories \cite{Dolanetal1999}. However, no such geometry
 is visible for interburst intervals data from epileptic hippocampal slices \cite{Schiffetal1994,Slutzkyetal2001}. 
Traditional surrogate analysis to assess statistical significance of UPOs in such return 
maps have provided mixed results \cite{Feietal2012,Quyenetal1997,Biswal-Dasgupta2002a}. 
New surrogate method that preserves the attractor shape in the shuffled surrogates have been found to be more 
suitable in such time series analysis for UPO detection \cite{Dolanetal1999}.  Even if the
apparent recurrent UPO like trajectories truly originate from an
underlying deterministic dynamics, how reliable are the fixed point,
stable and unstable manifolds derived from this small number of short
sequences from these noisy return maps with no apparent geometry? The
lack of geometry could also be due to noise and it is claimed that
presence of chaos can still be concluded if the observed UPOs are
recurrent and statistically significant. However, recurrent UPOs with
similar statistical significance have been reported from IBIs from
stochastic neuron models as well \cite{Christini-Collins1995}. Analysis of biological time series
and EEG data for seizure like activity and its suppression have been
done using tools other than UPO analysis also \cite{Geltrudeetal2012,Iasemidis-Sackellates1996,Gaoetal2012a}.

A clear distinction based on the underlying dynamics must exist
between the periodic dynamics of a system under chaos control and one
induced by periodic pacing or demand pacing.  The response of a brain
slice to external stimulation is largely unknown and different from a
physical systems.  Further, frequent external stimulation can cause permanent change
in the underlying dynamics or in the system parameters. This is evident from
a recent experiment showing control of seizures through electrical
stimuli \cite{Tang-Durand2012}. Therefore, it is necessary to show that the periodic dynamics under
chaos control is not due to any kind of biological synchronization
with the external stimuli.

Although the original chaos control experiment \cite{Schiffetal1994} assumed recurrent UPOs around the same trajectory as 
signature of determinism and chaos control was based on equation of manifolds derived from multiple UPO trajectories in 
the return map of IBIs, subsequent surrogate analysis have concluded presence of statistically significant 
"non-stationarity or drifting" UPOs from rat hippocampal slices \cite{Soetal1997, Soetal1998}. To ensure 
successful chaos control, even around such drifting UPOs, adaptive chaos control techniques 
have been proposed \cite{Inetal1995b}. Adaptive chaos control using more stringent UPO detection technique have been 
proposed \cite{Christini-Kaplan2000} and applied successfully in a repeat chaos control experiment in epileptic hippocampal slices \cite{Slutzkyetal2003}.
 
The experiment of Slutzky {\it {et al.}} \cite{Slutzkyetal2003} provides valuable new insight
and addresses some of these issues afresh, but a number of questions
still remain. In this paper,we attempt to answer some of 
these questions by computer simulation of this adaptive
chaos control experiment on a stochastic neural network model for the
epileptic brain slices. The
general question of deterministic chaos in neural systems is a vast
field of research and this work focuses only on the reliability of UPO
based chaos control applications in epileptic brain slices and the
conclusions drawn from them. In the concluding section we summarize
the main unresolved questions that prevent us from unambiguously
establishing that the bursting in epileptic brain slices is chaotic
dynamics in presence of noise rather than stochastic in origin.

\section{ Determinism from UPOs }
\label{sec:upo}

According to the cyclic theory of Chaos, chaotic dynamics is built upon "skeleton"
formed by UPOs \cite{Cvitanovic1988}. Due to ergodicity, the system
visits these UPOs along stable manifold and departs away from it along unstable manifold. 
However, time series analysis methods that chose the detection of these
statistically significant UPOs as a criterion for establishing deterministic chaos are 
more recent. A number of different techniques for search of UPOs in time series obtained from both low dimensional and high dimensional systems 
have been proposed \cite{Schiffetal1994,Soetal1997,Christini-Kaplan2000,Pierson-Moss1995,Pei-Moss1996,Gaoetal2009,Maetal2013} and subsequently applied to time series from physiological and 
biological systems \cite{Schiffetal1994,Iasemidis-Sackellates1996,Jiaetal2012,Braunetal2011,Quyenetal1997,Slutzkyetal2003,Gaoetal2012b,Yangetal2009,Pei-Moss1996,Used-Martin2013}. Most of these methods adopt surrogate methods
 to establish the statistical significance of the detected UPOs.

In an earlier work \cite{Biswal-Dasgupta2002b}, the topological recurrence method \cite{Pei-Moss1996}
was found to unambiguously reject short stochastic time series as non-chaotic when
other methods erroneously found statistically significant UPOs in them. All methods tend to accurately
discriminate chaotic and non-chaotic time series when data sets were sufficiently large (datasets of size $>4096$). 
However, results from shorter data sets were not reliable. Among various UPO detection methods, the 
topological recurrence method is the simplest and used regularly in analyzing biological time series although 
length of biological time series are usually small. Therefore, a rigorous analysis of the reliability of this geometric 
method in the context of length of data set, level of noise, nature of geometry and nonstationarity is necessary 
because each of these affects the geometry of the return map and, therefore, likely to affect the results.

\subsection{Characterizing UPOs}

The topological recurrence method for detection of UPOs adopts three 
levels of filtering to encounter UPO trajectories from the return map of a dynamics \cite{Pierson-Moss1995,Pei-Moss1996,Dolan2001b}.

\begin{itemize}

\item {\it Level} 0: Out of five sequential points, first three approach the identity line with decreasing perpendicular distance 
and the last three depart from the identity line with increasing perpendicular distance.
\item {\it Level} 1: Linear least square fit to the first three points must have a slope in the range ${|\lambda_s|}<1$
identifying a stable manifold and linear least square fit to the last three points must have a slope in the range $-1\ge{\lambda_{us}}>-\infty$
identifying an unstable manifold.
\item {\it Level} 2: Distance of the fixed point (estimated from the intersection point of the two straight line fits in the previous step)
 from the identity line must be less than or equal to $\frac{1}{m}$ times the average perpendicular 
distance of the five data points from the identity line.

\end{itemize}

Although {\it Level} 0 and {\it Level} 1 are based on the defining geometric properties of the UPOs, 
the {\it Level} 2 criterion was ad-hoc, 
and has been used differently in different applications \cite{Used-Martin2013,Brandt-Chen1996}. 
The {\it Level} 1 criterion in \cite{Pei-Moss1996} only included the slope 
condition for "flip" saddles. If the linear fits in {\it Level} 1 truly 
represent the stable and unstable manifolds, their intersection would lie 
on the identity line at the fixed point. In noisy data sets this condition may 
not be met exactly. Therefore,  in {\it Level} 2, a relaxed criterion 
is adopted to discriminate candidate UPO trajectories from the spurious ones. We have generalized the  {\it Level} 2 criterion to
arbitrary m, where as the original criterion \cite{Pei-Moss1996} was defined for $m=2$. Higher $m$ would demand the
estimated fixed point (the intersection point) to be more closer to the identity line. Although this is  expected 
to give more accurate UPO parameters, there will be lesser number of candidate UPO trajectories.  

\subsection{Statistical significance of UPOs}
To determine whether the detected UPOs actually originate from deterministic and not from stochastic dynamics, 
surrogate analysis methods have been proposed \cite{Soetal1997,Dolanetal1999,Dolan2001a,Pierson-Moss1995,Pei-Moss1996} . In the topological recurrence method \cite{Pei-Moss1996}, the $K$ value 

\begin{table*}
\centering
\begin{tabular}{|c|c|c|c|c|}
\hline
Noise & Level  & $\lambda_s$ & $\lambda_u$ & $x^*$ \\
\hline
& & mode \hspace{0.2cm}  range & mode \hspace{0.2cm}  range & mode \hspace{0.2cm}  range \\
\hline
& 0 & 0.2576\hspace{0.2cm}  1.2572 & -1.7850\hspace{0.2cm}  1.7006 & 0.6701\hspace{0.2cm}  0.6752 \\
0.0 & 1 & 0.2576\hspace{0.2cm}  1.2572 & -1.7850 \hspace{0.2cm} 0.8289 & 0.6747 \hspace{0.2cm} 0.5906 \\
& 2 (m$=$2) & 0.2548 \hspace{0.2cm} 1.2201 & -1.7950 \hspace{0.2cm} 0.8276 & 0.6724 \hspace{0.2cm} 0.4908 \\
& 2 (m$=$5) & 0.2560 \hspace{0.2cm} 1.2189 & -1.7986 \hspace{0.2cm} 0.8206 & 0.6730 \hspace{0.2cm} 0.4893 \\
& 2 (m$=$10) & 0.2550 \hspace{0.2cm} 1.2198 & -1.7974 \hspace{0.2cm} 0.7993 & 0.6736 \hspace{0.2cm} 0.4294 \\
\hline	

& 0 & -0.5533 \hspace{0.2cm} 1.4589 & -1.7798\hspace{0.2cm}  12.089 & 0.6556 \hspace{0.2cm} 1.5305 \\
0.1 & 1 & -0.5533 \hspace{0.2cm} 1.4589 & -1.7798\hspace{0.2cm}  6.3346 & 0.6556 \hspace{0.2cm} 1.5305 \\
& 2 (m$=$2) & 0.2271 \hspace{0.2cm} 1.4584 & -1.7998 \hspace{0.2cm} 6.3346 & 0.6591 \hspace{0.2cm} 0.5071 \\
& 2 (m$=$5) & 0.2374 \hspace{0.2cm} 1.4336 & -1.800 \hspace{0.2cm} 3.6838 & 0.6618 \hspace{0.2cm} 0.5145 \\
& 2 (m$=$10) & 0.2397 \hspace{0.2cm} 1.3819 & -1.7751 \hspace{0.2cm} 6.3288 & 0.6710 \hspace{0.2cm} 0.5024 \\
\hline	
& 0 & -0.5588\hspace{0.2cm}  1.5661 & -1.7832 \hspace{0.2cm} 13.7858 & 0.6522 \hspace{0.2cm} 1.5507 \\
0.2 & 1 & -0.5688 \hspace{0.2cm} 1.5097 & -1.7832 \hspace{0.2cm} 10.178 & 0.6521 \hspace{0.2cm} 1.5507 \\
& 2 (m$=$2) & 0.2312 \hspace{0.2cm} 1.5097 & -1.7832 \hspace{0.2cm} 10.173 & 0.6544 \hspace{0.2cm} 0.5658 \\
& 2 (m$=$5) & 0.2380 \hspace{0.2cm} 1.4797 & -1.8197 \hspace{0.2cm} 4.3313 & 0.6700 \hspace{0.2cm} 0.5263 \\
& 2 (m$=$10) & 0.2490 \hspace{0.2cm} 1.4092 & -1.7850 \hspace{0.2cm} 8.5878 & 0.6755 \hspace{0.2cm} 0.5122 \\
\hline	
\end{tabular}
\caption{Manifolds and fixed point estimate using topological recurrence criteria from 50 ensembles  $4096$ point dataset of H$\acute{e}$non map (Actual $\lambda_s$, $\lambda_u$, 
$x^*$ are 0.178, -1.69 and 0.6667 respectively.) }
\end{table*}

\begin{equation}
K = \frac{(N- \bar{N_s})}{\sigma_s},
\label{K_statistics}
\end{equation}

\noindent defines the statistical significance of the observed UPOs. Here N is the number of successful 
UPO encounters in the original data, $\bar{N_s}$ is the average number of successful UPO encounters in the  
surrogate data set, and $\sigma_s$ is the standard deviation of the number of observed UPOs in the 
surrogate data. $K >3$ signifies a statistical confidence level of more than 99$\%$ in rejecting the 
null hypothesis that the data originated from a random process \cite{Pei-Moss1996}.

\begin{figure*}
\centering
\includegraphics[width=12cm]{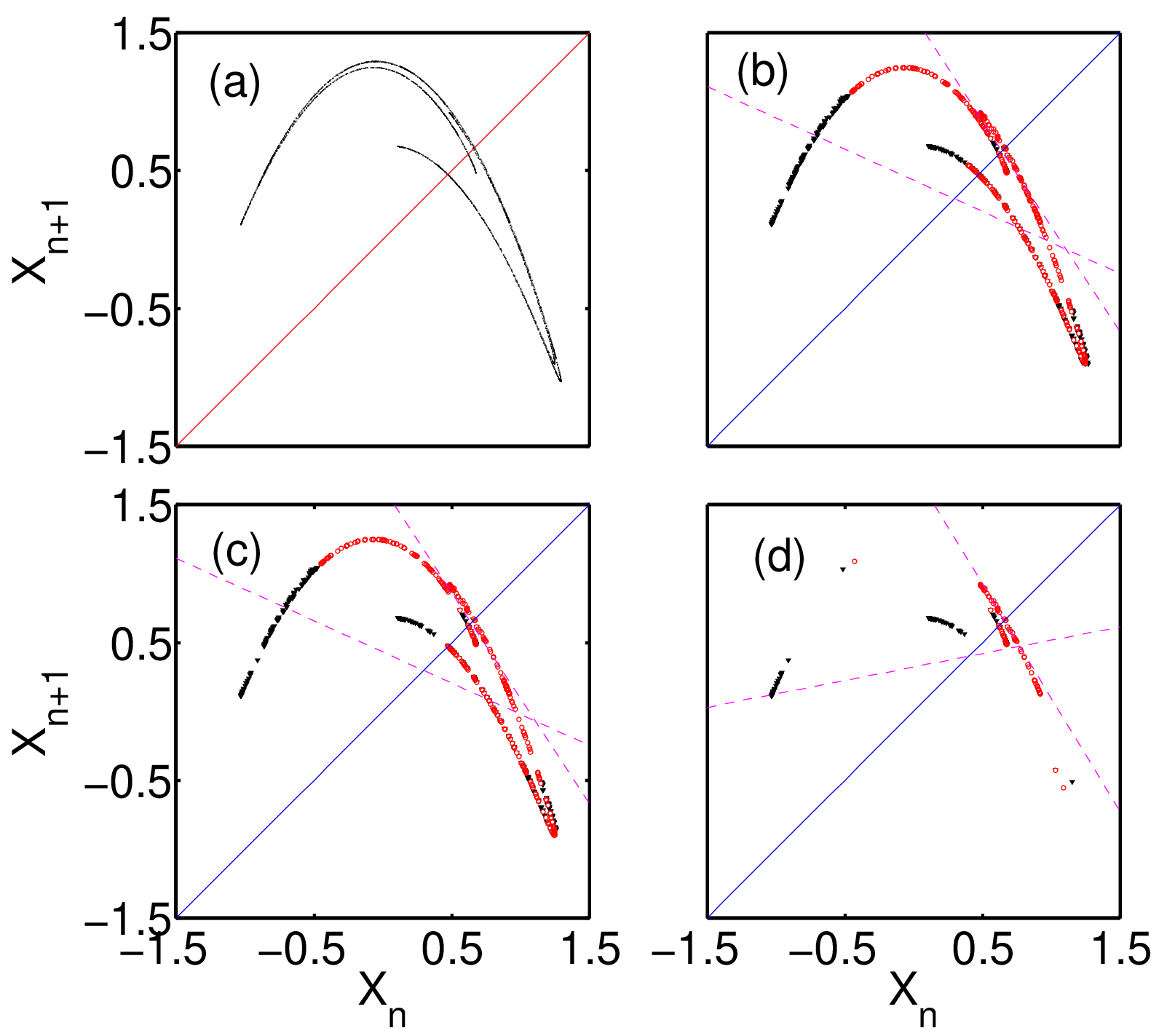}
\caption{UPO analysis of H$\acute{e}$non map ($X_{n+1}$ vs. $X_n$) without noise using topological recurrence criterion. 
(a) Actual return map, (b) {\it Level} 0 filtered, (c) {\it Level} 1 filtered and (d) {\it Level} 2 filtered UPO trajectories 
respectively.}
\label{fig1}
\end{figure*}

\begin{figure*}
\centering
\includegraphics[width=12cm]{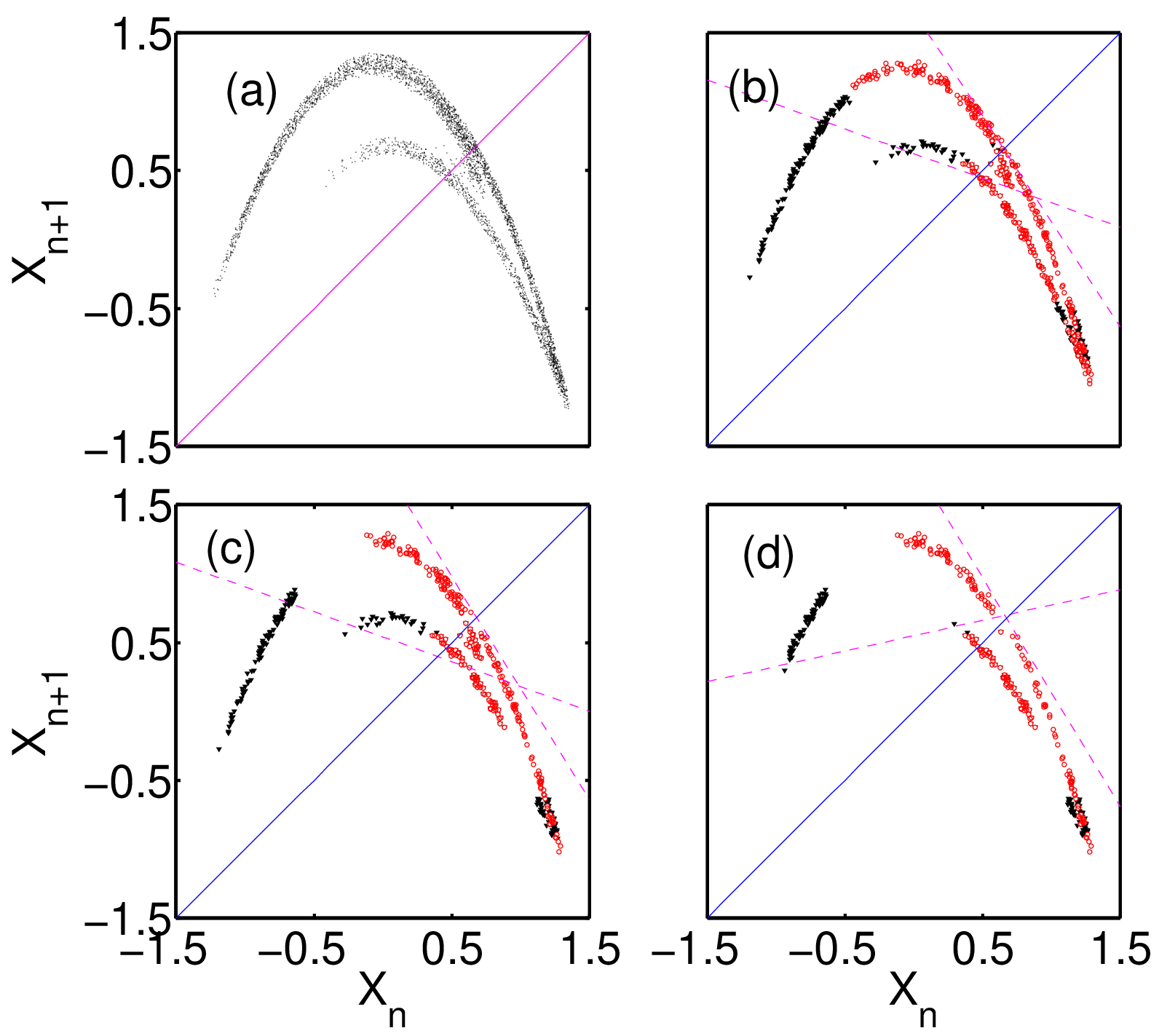}
\caption{UPO analysis of H$\acute{e}$non map ($X_{n+1}$ vs. $X_n$) with noise ($\epsilon=0.2$) using topological recurrence 
criterion. (a) Actual return map, (b) {\it Level} 0 filtered, (c) {\it Level} 1 filtered and (d) {\it Level} 2 filtered UPO trajectories 
respectively.}
\label{fig2}
\end{figure*} 

\begin{figure}
\centering
\includegraphics[width=8cm]{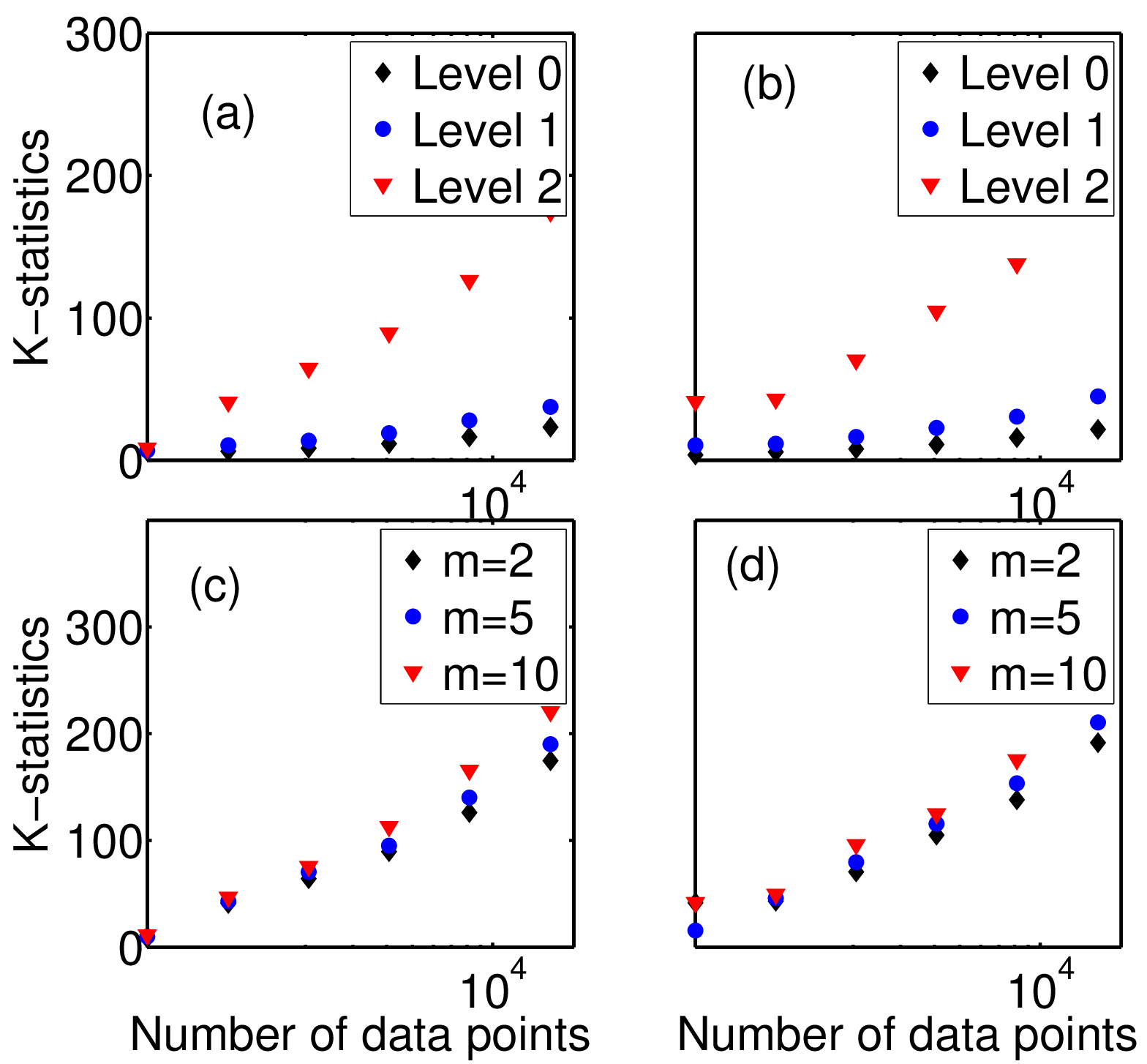}
\caption{Statistical significance of UPO analysis of H$\acute{e}$non map ($K$ vs. $n$) using topological recurrence 
criterion: ((a) without noise, (b) with noise $\epsilon=0.2$ for different levels of filtering, (c) {\it Level} 2 statistics for different value of m without noise $\epsilon=0.0$, (d) {\it Level} 2 statistics for different values of m with noise $\epsilon=0.2$).}
\label{fig3}
\end{figure} 

\subsection{Results}
We applied the topological recurrence method to estimate dynamical parameters
by looking for candidate UPOs in the first return map of three known deterministic chaotic systems, each having 
a different geometry. We then compared the estimated UPO characteristics with the true values. Dependence on the
effect of noise, length of data sets and nature of geometry on the estimation of dynamical parameters are 
studied in detail.  For each data set, at each {\it level} of filtering, a number of candidate UPOs are obtained.
For each of these trajectories, the three parameters characterizing the UPO trajectories ($\lambda_s$, 
$\lambda_{us}$, and the fixed point) are determined. Statistical fluctuations of the estimated dynamical parameters are
analyzed by measuring the {\it mode} and the {\it range} (difference between maximum and minimum values). This is done for
different levels of noise, different lengths of data sets, and different levels of filtering.\\

\noindent The three systems being studied are : H$\acute{e}$non map, Ikeda map and R$\ddot{o}$ssler attractor. 
The dynamics of H$\acute{e}$non map is described by 
\begin{eqnarray}
X_{n+1} &=& 1 + bY_n - aX{^{2}_n}, \nonumber \\
Y_{n+1} &=& X_n + \epsilon Z, 
\label{Henon}
\end{eqnarray}

\noindent where a$=$1.2, b$=$0.3, and $\epsilon$ denotes strength of the dynamical noise. Z is a
uniform random variable in the range [-1,1]. For these set of parameter values and $\epsilon=0$, the true 
values characterizing the UPOs are:  $\lambda_s=0.178$, $\lambda_{us}=-1.69$, and fixed point at $X^*=0.6667$ \cite{Dolan2001b}. 

H$\acute{e}$non map has been analyzed earlier using only the {\it Level} 0 topological recurrence method combined with a 
matrix fit algorithm \cite{Dolan2001b}. The dynamical parameters were estimated and analyzed with or without 
noise. In this work we try to understand the deviation of estimated results from actual values in the presence of noise, 
its dependence on length of data sets as well as its dependence on the nature of the geometry. At each {\it Level},  
a number of candidate UPO trajectories are observed and dynamical parameters are computed from linear fits to 
these trajectories.  These results are summarized in Table 1. 

In absence of noise, {\it mode} of the estimated dynamical parameters is nearly the same for different levels of filtering. 
This has been observed earlier also \cite{Dolan2001b}.
 In absence of noise, {\it Level} 0 is adequate for the purpose of establishing statistically significant UPO detection.
{\it Range} of the estimated parameter values decreases as the filtering criteria is made more stringent, i.e., as one goes from 
{\it Level} 0 to {\it Level} 2. In presence of noise, the geometry of the return map also gets noisy. As expected,
this leads to inaccurate estimate as well as increase in the values of {\it range}. The value of the {\it mode} for the stable 
manifold estimates for {\it Level} 0 and 
{\it Level} 1 filtering criteria deviates significantly from the {\it mode} of {\it Level} 2 estimated values, although the secondary peak is 
nearly the same as the {\it mode} of {\it Level} 2 estimate. 
This signifies occassional failure of {\it Level} 0 and {\it Level} 1 criteria in the presence of noise. Also, the {\it mode} of 
stable manifold estimates is nearly 40-50 $\%$ more than the true value, while the deviations in the {\it mode} of 
unstable manifolds and fixed point estimates are less than 10$\%$. We have added a further filter in {\it Level} 2, i.e., 
the value of $m$. The results presented in Fig. \ref{fig1,fig2} are for $m=2$ and those summarized in Table 1 are for $m=2$, $5$ and $10$. 
Increasing $m$ does not lead to significant improvement in the UPO parameters, 
but the number of candidate UPO trajectories decreases significantly.  With increasing m,
 K-value remains nearly identical for small datasets and increases for large datasets with or without noise as shown in Fig. \ref{fig3}.

Surprisingly, while the variations in UPO parameter estimates increase with noise, no significant 
change is observed in the K-statistics shown in Fig. \ref{fig3}. This means that the topological recurrence method is fairly robust in establishing
determinism in the presence of moderate noise in data sets of moderate length (data size $>1024$). For the same length data set, the number of candidate 
UPOs detected at each level decreases with increase in strength of noise. Unchanged K-statistics indicates that
the number of candidate UPOs detected in the surrogate data sets also decreases.    

To understand how the topological recurrence method depends on the type of geometry in the return map of the dynamical system, two
other systems are also analyzed. The Ikeda map is described by 
\begin{eqnarray}
t_n &=& a - b/(1+X_n^2+Y_n^2), \nonumber \\ 
X_{n+1} &=& 1 - uX_n \text{cos}(t_n) - uY_n  \text{sin}(t_n)+\epsilon z_1, \nonumber \\ 
Y_{n+1} &=&  uX_n \text{sin}(t_n) + uY_n  \text{cos}(t_n)+\epsilon z_2 ,
\label{Ikeda}
\end{eqnarray}

\noindent where $a=0.4, b=6.0, u=0.9$. For these set of parameter values, the fixed point is 
$1.0550$, the slope of the stable manifold is $0.6634$ and the slope of the unstable 
manifold is $-1.221$ \cite{Skufca-Bollt2003}. $\epsilon$ denotes strength of dynamical noise, $z_1$ 
and $z_2$ are uniform random variables in the range [-1,1]. The data for $X$ are analyzed for
UPOs using topological recurrence criteria. In absence of noise, the {\it mode} of all the three 
estimated parameters improve significantly as the filtering criteria is made more stringent. However, 
stable manifold estimate deviates significantly from the actual value even for {\it Level} 2 filtering 
({\it mode} is $0.070592$ and {\it range} is $0.6930$ for $50$ datasets of length $4096$). In 
absence of noise, out of $50$ data sets of size $4096$ each, only $35$ give successful UPO 
candidates at {\it Level} 2, out of which $17$ data sets give just one successful UPO candidate. 
In presence of noise, the fixed point estimate deviate from true value by $10\%$ for {\it Level} 2 
filtering (for $\epsilon = 0.2$). Unstable manifold estimates deviate by 20 $\%$ for different levels of 
filtering with or without noise. With addition of noise, there is uniform increase in the {\it range} of all 
estimates indicating greater deviations from actual parameter values. When dynamical noise is introduced, 
$42$ data sets give successful UPO candidate at {\it Level} 2 out of which $15$ data sets have just one 
successful candidate UPO. This indicates that for this geometry, the system dynamics do not visit close to the 
fixed point frequently and the estimated parameters deviate significantly from true values because of
this reason.

The R$\ddot{o}$ssler attractor is a continuous time dynamical system described by 
\begin{eqnarray}
\dot{x} &=& -(y+z), \nonumber \\
\dot{y} &=& x+ay, \nonumber \\
\dot{z} &=& b+ z(x-c)+\sqrt{2D}\Xi(t),
\label{Rossler}
\end{eqnarray}

\noindent where $\Xi(t)$ is gaussian white noise with zero mean and unit variance. 
D is the amplitude of dynamical noise. The values of the parameters are $a=0.2$, $b=0.2$, and
$c=5.7$.  From X=0 plane Poincar$\acute{e}$ section is obtained and the variable $y$ is used for 
further analysis \cite{Dolan2001b}. The actual fixed point, stable manifold and unstable manifold are 
$-8.39$, $-2.34\times$10$^{-7}$ and $-2.40$ respectively.

For this system, the topological recurrence criteria shows strong dependence on the geometry of
the Poincar$\acute{e}$ section. In absence of noise, the {\it mode} of the estimates of $\lambda_s$ 
vary with each level of filter ($-0.571$ with a {\it range} of $0.9649$ for both {\it Level} 0 and 
{\it Level} 1, and $-0.025129$ with a {\it range} of $0.58709$ for {\it Level} 2). These estimates 
are significantly different from the actual values. It appears that the method fails because $\lambda_s$
is close to zero. Unstable manifold estimates for {\it Level} 0 and {\it Level} 1 filtering are close to 
actual value ({\it mode} is $-2.1133$ and {\it range} is nearly $2.0$), but deviates significantly from 
the actual value for {\it Level} 2 ({\it mode} is $-1.1049$ with {\it range} $0.6406$). Fixed point estimates 
are fairly accurate for {\it Level} 2 filtering criteria ({\it mode} is $-8.4314$ with {\it range} of $0.94955$).
For this geometry, the estimates do not deteriorate much with introduction of noise, although the variability
increases marginally. In absence of noise, out of $50$ data sets of size $4096$, $16$ data sets give successful 
UPO candidates at {\it Level} 2 filtering, out of which $13$ data sets have just one successful candidate UPO, the
reason for large deviation in stable and unstable manifold. Similar results are obtained when dynamical noise 
is introduced in the system. Above observations show that geometrical structure greatly influences the frequency 
of UPO visit as well as the accuracy in the estimation of dynamical parameters. However, the statistical significance 
of the UPOs assessed at all levels of noise is above $95\%$. The 
topological recurrence method is robust in establishing all these systems to be deterministic even in the presence 
of moderate noise. Large values of m signifies close visit to UPOs. For H$\acute{e}$non map, UPO candidates
 satisfying conditions for higher m are easily obtained, while for Ikeda map and R$\ddot{o}$ssler attractor, 
even UPO candidates satisfying $m=2$ criteria is not found frequently. This means that choice of m value 
to improve UPO detection will depend on the system under investigation.  

In the above analysis, random shuffled surrogates were used. It has been pointed out that random
surrogates that preserve the attractor shape are more appropriate \cite{Dolanetal1999}. 
From, the results of the surrogate analysis of H$\acute{e}$non map using surrogates that preserve attractor shape,
 it is observed that K-statistics for {\it Level} 0 is similar to that of shuffled surrogates for different strength of noise 
and length of data sets. No candidate UPO was found for {\it Level} 1 and {\it Level} 2 for data sets of length less than 16384. 
For data set of length 16384 and m $=$ 2,  {\it Level} 1 and {\it Level} 2 statistical significance is $>$ 99 $\%$. As noise is 
added, K-statistics value decreases for {\it Level} 1 and {\it Level} 2. Similar results are obtained for m $=$ 5  and 
m $=$ 10. As m is increased, K-value for {\it Level} 0 increases while it decreases for {\it Level} 1 and {\it Level} 2.
We observe that for systems with deterministic geometry, the random surrogates that preserve attractor shape are less 
likely to have candidate UPOs, resulting in higher K-statistics. However, if the return map does not have any visible 
geometry, the attractor shape preserving surrogates do not improve statistical tests significantly.

These results suggest that the UPO parameter estimates are good when the system dynamics repeatedly 
visits very close to the UPOs. With increase in the length of data sets, number of visits as well as number of 
close encounters with UPOs increase yielding very good estimates of the fixed point and slopes of the
stable and unstable manifolds. Close encounter with UPOs and accuracy of dynamical parameter estimation 
depends strongly on geometrical structure. The statistical significance of UPO detection is above $95\%$ for the 
diverse cases examined indicating this to be a general result. As noise level is increased, smearing of the UPO 
structure takes place and manifold estimates deteriorate, although statistical significance remains high. With 
small length of data sets and high level of noise, obtaining statistically significant result depends on the frequency 
of UPO visits, which in turn depends on the geometry of the return map. There is a decrease in number of successful 
encounters at each level. We also observe that the UPO parameters  do not improve significantly by increasing $m$. 
Increasing $m$ is analogous to looking for close encounters to UPOs, which results in decrease in number of 
successful encounters. Even at low level of noise and for long data sets, the estimated UPO parameter 
values are not accurate enough for control applications. Another crucial observation is that the method works
well for the dynamics where the third point of the candidate UPOs is actually close to the fixed point. The 
topological recurrence criteria assumes that this point belongs to both the stable and the unstable manifold.

\subsection{Modified UPOs selection criterion using dynamical transformation method}
The results and analysis presented in the previous section point to two major shortcomings of the topological 
recurrence criteria. The fixed point estimate as determined from the intersection of the linear fits to the approaching
and receding set of points is inaccurate. The requirement that the fixed point need to be close to the identity
line rather than a specific point on the identity line introduces additional error. 

Instead of using least square fit, we adopt a method used by Dolan \cite{Dolan2001b} to obtain estimates of fixed point 
and manifolds. In the vicinity of UPO, system dynamics is approximated as 
\begin{equation}
X_{n+1} = AX_n + BX_{n-1}+C.
\label{matrixfit}
\end {equation}
The {\it Level } 0 criterion gives five consecutive points defining the UPO trajectory near the fixed point.  
Slopes of the manifolds and the fixed point of the detected UPOs were obtained by solving the three simultaneous 
linear equations obtained from the three triplets of the candidate UPO trajectory \cite{Dolan2001b}. This gives significantly improved
result than the least square method of {\it Level} 1.  
Instead of the average distance from the identity line criterion in {\it Level} 2, we use a new criterion that uses the 
So {\it et al.} \cite{Soetal1997} method of dynamical transformation that gives very good estimate of period $1$ 
fixed point even in presence of noise.    
We redefine the {\it Level } 2 criterion as  

\begin{equation}
{\it Level}~2:~~~~~~\frac{|x_2^\star - x_1^\star|}{x_2^\star} <= \delta,
\end{equation}
where $x_2^\star$ is the fixed point estimate from dynamical transformation method \cite{Soetal1997}, 
and $x_1^\star$ is the fixed point estimate from {\it Level} 1 using Eqn. \ref{matrixfit}. $\delta$ is a measure of close encounter 
with UPOs.

\begin{table*}
\centering
\begin{tabular}{|c|c|c|c|c|}
\hline
Noise & Level  & $\lambda_s$ & $\lambda_u$ & $x^*$ \\
\hline
& & mode \hspace{0.2cm}  range & mode \hspace{0.2cm}  range & mode \hspace{0.2cm}  range \\
\hline
& 0 & 0.1064\hspace{0.2cm}  0.8919 & -1.7929\hspace{0.2cm} 0.8361 & 0.6467\hspace{0.2cm}  0.3764 \\
0.0 & 1 & 0.1064\hspace{0.2cm}  0.8919 & -1.7929\hspace{0.2cm} 0.8361 & 0.6467\hspace{0.2cm}  0.3764 \\
& 2 & 0.1698 \hspace{0.2cm} 0.0562 & -1.7740 \hspace{0.2cm} 0.0656 & 0.6709 \hspace{0.2cm} 0.0012 \\
\hline	

& 0 & 0.1428 \hspace{0.2cm} 7.6847 & -1.7719\hspace{0.2cm}  10.417 & 0.6593 \hspace{0.2cm} 0.4338 \\
0.1 & 1 & 0.1428 \hspace{0.2cm} 1.724 & -1.7719\hspace{0.2cm}  9.4743 & 0.6593 \hspace{0.2cm} 0.4338 \\
& 2 & 0.1867 \hspace{0.2cm} 0.6547 & -1.8431 \hspace{0.2cm} 0.8908 & 0.67014 \hspace{0.2cm} 0.0013 \\
\hline	
& 0 & 0.1432\hspace{0.2cm}  12.766 & -1.7008 \hspace{0.2cm} 20.417 & 0.6512 \hspace{0.2cm} 1.8558 \\
0.2 & 1 & 0.1294 \hspace{0.2cm} 1.9708 & -1.7008 \hspace{0.2cm} 18.931 & 0.6512 \hspace{0.2cm} 1.8558 \\
& 2 & 0.1832 \hspace{0.2cm} 0.9965 & -1.9436 \hspace{0.2cm} 1.5517 & 0.6696 \hspace{0.2cm} 0.0013 \\
\hline	
\end{tabular}
\caption{Manifolds and fixed point estimate using modified Pei-Moss criterion from 50 ensembles  $4096$ point data set of H$\acute{e}$non map (Actual $\lambda_s$, $\lambda_u$, 
$x^*$ are 0.178, -1.69 and 0.6667 respectively.) }
\end{table*}

The results from this hybrid method that combines the original topological recurrence criterion \cite{Pei-Moss1996} with the 
matrix fit algorithm \cite{Dolan2001b} and dynamical transformation method \cite{Soetal1997} for the H$\acute{e}$non map are 
presented in Table 2. Adopting the matrix fit algorithm Eqn. \ref{matrixfit} gives different values for the fixed point 
and the slopes of the manifolds than the original topological recurrence criterion. There is significant improvement 
in the accuracy of the UPO parameters compared to the results from the topological recurrence criterion summarized 
in Table 1. 

Values of $\delta$ used for H$\acute{e}$non map are  $0.001$, $0.005$ and $0.005$ for the noise values $0.00$, $0.10$ 
and $0.20$ respectively. With or without noise, estimates approach the actual values as filtering is made more stringent. 
Estimates of stable manifolds deviate by 5$\%$ to 10$\%$ as noise varies from $0.0$ to $0.2$. The estimates of the unstable
manifold deviate from 6$\%$ to 15$\%$ for the same noise variation. At {\it Level} 2, the fixed point estimates are accurate 
up to $0.5\%$ for these noise values. For all the three parameter estimates, results for {\it Level} 0 and {\it Level} 1 are 
identical for noise level $0.0$ and $0.1$, while for noise $0.2$, there is marginal variation only in the stable manifold estimate. 

Values of $\delta$ used for R$\ddot{o}$ssler attractor are  $0.001$, $0.005$ and $0.005$ for the noise values $0.00$, $0.10$ and 
$0.20$ respectively. Results for the stable manifold estimates at {\it Level} 2 are $-0.01648$, $-0.02275$, $-0.01989$ for 
$D=0.0,0.1,0.2$ respectively, showing large deviation from the true value. This is due to the small magnitude of the slope of the 
actual stable manifold. Results for the unstable manifold and fixed points, for different levels of filtering and strength of noises, 
are identical with accuracy up to $12\%$ and $1\%$ respectively. Results for {\it Level} 0 and {\it Level} 1 are identical 
for strength of noise varying from $0.0$ to $0.2$ for all the three parameter estimates.

For Ikeda map, value of $\delta=0.1$ is used for three noise values $0.00$, $0.10$ and $0.20$. For {\it Level} 2 filtering,
fixed point estimates deviate up to $3\%$ for different strengths of noise. Result for unstable manifold differs by $50\%$
for noise $0.0$ and $0.1$, and by $70\%$ for noise $0.2$. Estimates of stable manifold differs by $50\%$ for {\it Level} 2 
filtering (for different strengths of noise). Results of {\it Level} 0 and {\it Level} 1 are not identical as is observed
in case of H$\acute{e}$non map and R$\ddot{o}$ssler attractor.

The results at {\it Level} 0 and {\it Level} 1 filter are not close to actual UPO parameters for the three systems studied. 
Otherwise, UPO parameters at the end of {\it Level 2} of the hybrid method are very close to the actual values except for 
the Ikeda map. Comparing the results from the three systems analyzed in this work, we find that with matrix fit algorithm, 
only a very small fraction of the UPOs detected in {\it Level} 0 are rejected at the next level of filter {\it Level} 1. The 
{\it Level} 2 filter is more stringent and is the reason behind very accurate parameter estimation even in the presence of 
moderate noise. This hybrid method works better than the topological recurrence method. Better parameter estimation results from
adoption of a more accurate fixed point in {\it Level} 2. Close encounter with UPOs can be adjusted depending on the 
geometrical structure. If $\delta$ value is large, stable and unstable manifold estimate won't be good but fixed point estimate 
will be good, which is primarily due to the inclusion of the dynamical transformation method.    

\section{Chaos control using adaptive UPO tracking}
Results of previous section suggest that unambiguous detection of UPOs is often possible in short and noisy
biological time series through topological recurrence criteria. However, we also observe that
estimation of quantitative parameters characterizing the UPOs deteriorate rapidly with addition of noise.
In this context, it is important to reinterpret control strategies reported to be successful and carefully redesign such experiments
in future. This is why the deterministic interpretation of the original chaos control experiment on epileptic brain slices 
was questioned earlier \cite{Christini-Collins1995,Biswal-Dasgupta2002a}.

The results and analysis presented in the previous section show the unreliability of the UPO parameters 
obtained from return maps with noisy geometry even when the data is from known deterministic
dynamics. Return maps plotted from biological time series invariably lack any noticeable geometry \cite{Schiffetal1994,Slutzkyetal2003}. Although
chaos control is an efficient tool for confining system dynamics to any of the periodic orbits by suitable 
intervention at appropriate times, reliable estimation of fixed point, slope of manifolds and nonstationarity 
pose tough challenge for good chaos control. To understand the apparent success of the  repeat chaos control experiment on 
epileptic brain slices by Slutzky {\it et al.} \cite{Slutzkyetal2003,Slutzkyetal2002,Slutzkyetal2001}, it is necessary to first judge if the apparent success unambiguously originate from the underlying deterministic dynamics and made possible because of the adoption of more stringent detection and control strategies.
We first  examine the efficacy of the adaptive UPO tracking along with stable manifold perturbation (SMP) method that is used to obtain good control results for H$\acute{e}$non map (Eqn.\ref{Henon}) in a known deterministic system \cite{Christini-Kaplan2000}. 

In the vicinity of UPOs, system dynamics can be approximated by Eqn. \ref{matrixfit} which can be written as
\begin{equation}
\begin{split}
X_{n+1} = (\lambda_s+\lambda_u)X_n -\lambda_s\lambda_u~X_{n-1} \\ + X^{\star}(1+\lambda_s\lambda_u-\lambda_s-\lambda_u),
\end{split}
\label{lin2}
\end{equation}
where $\lambda_s$, $\lambda_u$ and $X^\star$ are actual stable eigenvalue, unstable eigenvalue and fixed point respectively \cite{Christini-Kaplan2000}.
Chaos control by stable manifold perturbation (SMP) is done through
\begin{equation}
 X_{n+1} = \lambda_s(X_n -X^{\star}) +X^{\star}.
\label{smp}
\end{equation}
UPO parameters estimated using the hybrid topological recurrence criterion is used for initiating SMP and then system dynamics is
allowed to evolve naturally. If the trajectories stay within a control region, no further perturbation is applied. SMP is again applied 
when the system comes out of the control region. To do this, first natural triplets ($X_{n+1}$, $X_n$, $X_{n-1}$), where $X_{n+1}$
correspond to natural dynamics, are obtained. Least square fit using singular value decomposition (SVD) is performed on a set 
of $10$ natural triplets to obtain new estimates of stable manifold and fixed point \cite{Christini-Kaplan2000}. SMP is again applied using revised estimated 
values. The process is continued till the control cannot be improved further. 

\begin{figure}
\centering
\includegraphics[width=8cm]{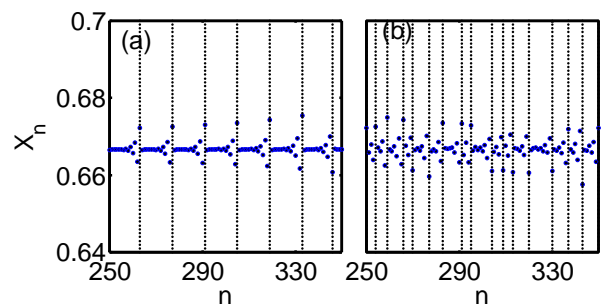}
\caption{ Adaptive control of H$\acute{e}$non map (a) without noise, (b) with noise=0.005}
\label{fig4}
\end{figure}

In the absence of noise and nonstationarity, our UPO parameter estimates improve with each iteration and approach the
actual values. With accurate UPO parameter estimates, the control lasts for significant duration without needing any external 
intervention as shown in Fig. \ref{fig4}. However, with the addition of noise, accuracy of the UPO parameter estimates deteriorates  and
adaptive tracking fails at regular intervals. With no noise, control application was needed on an average every $14$th iteration while with 
a small value of noise, control application was needed every $7$th iteration on an average. This signifies the difficulty of adaptively controlling a known deterministic system even in the presence of very low level of dynamical noise that does not alter the 
geometry of the return map appreciably \cite{Channelletal2009}.

To re-examine the apparent success of the repeat chaos control
experiment on epileptic brain slices using the stringent adaptive tracking method 
\cite{Slutzkyetal2002,Slutzkyetal2001}, we repeat the procedure on a stochastic neural network model of epileptic 
brain slices \cite{Biswal-Dasgupta2002a,Biswal-Dasgupta2002b} with same strategy. 
For completeness, a brief description of this computer model for epileptic hippocampal slices
\cite{Mehtaetal1993} is presented below. The neural network consists of $N$
McCulloch-Pitts (binary) neurons $S_i\in\{0,1\}$, each excitatory,
where $0$ and $1$ represent the low and high firing rates
respectively. The inhibitory neurons are collectively modelled by a
background inhibition that is proportional to the number of active
excitatory neurons.  The ``local field'' at the $i$th neuron at time
$t$ is given by

\begin{eqnarray}
h_i(t) & = & \sum_{j=1}^{N}[\{J_{ij}S_j(t)-wS_j(t)\} \nonumber\\
       & ~ & +\lambda\{K_{ij}S_j(t-\tau)-wS_j(t-\tau)\}],
\label{localfield}
\end{eqnarray}

\noindent where $w$ is the relative strength of inhibition, $\lambda$
is the relative strength of the delayed signal, and $\tau$ is the time
delay associated with the delayed signal. The four terms from left,
fast global excitation, fast global inhibition, slow global excitation
and slow global inhibition are essential \cite{Traubetal1993} for a realistic
modelling of neural oscillations in the hippocampus.  The ``time'' $t$
is discrete, with each unit (referred to as one ``pass'')
corresponding to the updating of $N$ neurons. The neurons are updated
one at a time in a {\it random sequence}, according to the rule:
$S_i(t+1)=1$ if $h_i(t) \ge 0$ and $S_i(t+1)=0$ if $h_i(t) < 0$.

A fixed number $(q)$ of random low activity patterns (``memories'')
$\{ \xi ^\mu _i \},~ i=1,2,.....,N;~ \mu=1,2,.....,q$, are stored in
the synaptic matrices $J_{ij}$ and $K_{ij}$ as
$J_{ij} = \Theta(\sum_{\mu=1}^{q}\xi_i^\mu \xi_j^\mu)$ and
$K_{ij} = \Theta(\sum_{\mu=1}^{q}\xi_i^{\mu+1} \xi_j^\mu)$
respectively.  $\Theta(x)=1(0)~\mbox{if}~x>0(x\le0)$, $\xi_i^{q+1} =
\xi_i^1$ and $J_{ii}=K_{ii}=0$. Each memory contains n ``active''
neurons chosen randomly and $ n=\sum_{i=1}^N \xi_i^\mu <<N$ ensures
low average activity of the network in the absence of any external
stimulus. Due to the properties of the synapses and with appropriate choice
of other parameters (specifically, for $\lambda > 1$ and $w < 1$), the
network exhibits a low-activity limit cycle in which all the $q$
memories are visited sequentially \cite{ Mehtaetal1993}. In this work, 
the chosen parameter values are: $N=200,~ q=20,~ n=10,~
w=0.6,~ \lambda=2,~ \tau=2$ passes. 

The hippocampal brain slices used in the chaos control experiments
were chemically kindled to generate epilepsy. This process is 
replicated in this computer model by reduced inhibition (equivalent to
excitation) accompanied with hebbian learning \cite{Mehtaetal1993}. The network
was excited by chosing $w=0.24$ for $50$ initial passes during which 
new fast connections were created through a hebbian learning mechanism. 
Addition of these new connections generates spontaneous population 
bursting similar to the bistable dyanmics of epileptic bursts and it has
been established that in this computer model of epileptic brain slices,
the underlying dynamics of this epileptic bursts are stochastic in nature \cite{Biswal-Dasgupta2002b}.      

\begin{figure}
\centering
\includegraphics[width=8cm]{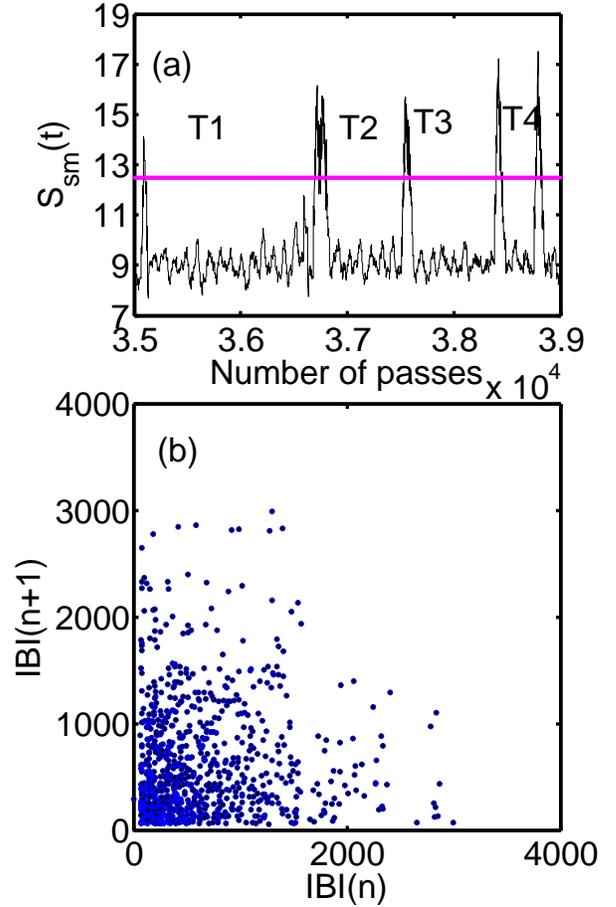}
\caption{Stochastic neural network showing epileptic bursts (a) Smoothened activity and (b) Return map}
\label{fig5}
\end{figure}

Although this model with binary neurons and synapses may be inadequate
for truly describing neurobiological dynamics of an hippocampal slice that 
depends crucially on the detailed biophysical properties of individual neurons, 
it incorporates a number of important neurobiological features
of kindled brain slices. Networks of binary neurons can still
qualitatively describe the collective dynamical behavior of biological
networks.  In the context of this paper, however, this
model is ideally suited for simulating the control experiments on
computer similar to the real experiments on brain slices.  Further,
the proven stochastic nature of the bursting dynamics
makes it a suitable surrogate model on which UPOs detection and adaptive chaos
control experiments should be reasonably difficult to reproduce with comparable 
success and qualitatively different from the real brain slices, if the bursting
dynamics of the real biological systems were deterministic in nature. Typical bursting dynamics 
and the first return maps of inter burst intervals (IBIs) from the computer model are shown in Fig. \ref{fig5} for 
completeness.

Fixed point of the dynamics is estimated from the first return map of the IBIs  using dynamical transformation method of So
{\it {et al.}} \cite{Soetal1997} and manifolds are estimated using modified UPO selection criterion. To apply control,
fixed radius($R_F$) within which control has to be applied is fixed as well as the desired radius 
within which IBIs will be left unperturbed, called control radius($R_C$). Control is switched on
when current IBI lies in $R_F$ and continued till the time it enters $R_C$ \cite{Slutzkyetal2003}. 
 Once we get inter-burst interval(IBI) within $R_F$, time of next IBI is calculated using
\begin{equation}
T_{n+1}=\lambda_s(T_n-fp)+fp,
\end{equation}
where $T_n$  is the time interval between $n^{th}$ and $(n-1)^{th}$ IBI, $\lambda_s$ 
is the slope of the stable manifold and $fp$ is the fixed point. 
If a burst is obtained before or up to the estimated time of burst, time for the  
next burst is calculated and the process is continued. Otherwise inhibition ($\omega$) is reduced for next five passes 
to generate a stimulated burst \cite{Biswal-Dasgupta2002b}. 

\begin{figure}
\centering
\includegraphics[width=8cm]{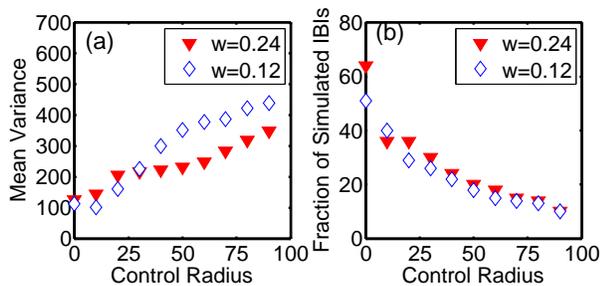}
\caption{ Effect of control radius $R_C$ on (a) Variation of variance and (b) fraction of stimulated IBIs }
\label{fig6}
\end{figure} 

\begin{figure*}
\centering
\includegraphics[width=15cm]{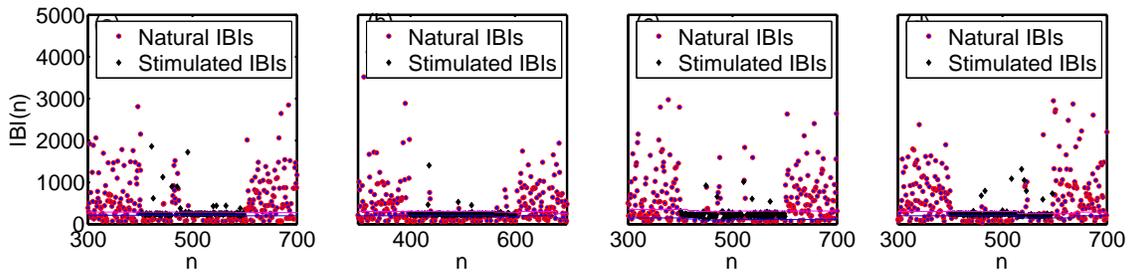}
\caption{{ Effect of control radius $R_C$ and inhibition strength(w) on the adaptive 
chaos control of stochastic neural network for epileptic brain slices. $R_{NT}$=40, FAM=50, $R_F$=200 (a) w=0.12, $R_C$=50,
(b) w=0.12, $R_C$=100, (c) w=0.12, $R_C$=50, (d) w=0.06, $R_C$=50 }}
\label{fig7} 
\end{figure*}

It is found that as $R_C$ is decreased, more and more number of control stimuli has to be applied and variance of IBIs decreases Fig. \ref{fig6}. 
As $R_F$ is decreased, lesser number of control stimuli has to be applied and variance of IBIs increases. 
To get quality control, both $R_C$ and $R_F$ value have to be optimized taking into consideration variance of IBIs. 
Quality of control is improved by adaptive tracking of UPOs. We use 10 natural triplets to refine the value of fixed point and manifolds
by least square fit method using singular value decomposition. Care is taken that, naturally terminated IBI must lie within a  
small radius ($R_{NT}$) from previous estimate of fixed point. To tackle the problem of fluctuations in the fixed point estimate, 
previous fixed point estimate is refined if and only if new estimate lies within a small radius from the previous fixed point called,
 a process known as fixed point adjustment maximum (FAM)\cite{Slutzkyetal2003}.  This process helps tackle the problem of nonstationarity.
Control applications for different values of the control radii are shown in Fig. \ref{fig7}. It is observed that application of chaos 
control reduces the concentration of IBIs over and above 
the fixed point. The SMP control ensures external stimuli if the IBIs are above the fixed point because slope of the stable manifold is usually small. Therefore, lesser the concentration above fixed point, better is the quality of control. 
Control achieved was good but deteriorated once adaptive tracking was introduced as there is modification
either to the stable eigenvalue or to the fixed point. Number of passes required to obtain desired number of
IBIs decreases as control is applied.  

As control radius ($R_C$) is increased, the fraction of IBIs above the 
fixed point decreases and the fraction of Stimulated IBIs increases. 
As the strength of inhibition is reduced, less number of IBIs are simulated keeping other parameters constant. 
Decreasing concentration of points over and above
fixed point requires more and more number of control stimuli, indicating  equivalence of chaos control and demand pacing.
Result of our adaptive chaos control application on the  stochastic neuron model of epileptic brain slice is shown in Fig. \ref{fig7}. 
Quality of control, effect of control radii on the variation of the variance and fraction of stimulated IBIs are found to be in very good agreement with the results of adaptive chaos control experiment performed on actual biological network \cite{Slutzkyetal2003}. 
This indicates that more stringent UPO detection or adaptive UPO control may not necessarily be the reason of successful chaos
control in this experiment. 
This comparison also indicates that the true system and the stochastic neural network models can be discriminated through UPO analysis 
using topological recurrence criteria, but, the two systems cannot be discriminated on the basis of the chaos control
although one is clearly a stochastic system.

\section{Discussion and Conclusion}
Nonlinear dynamics tools have been immensly useful in both the understanding of biological systems and  
specific control applications in such systems. However,  in biological systems, applications of nonlinear time 
series analysis tools for establishing determinism have not been straightforward. This work addresses two 
specific but important issues in this context - establishing determinism through detection of statistical significant 
UPO trajectories in the return map of interspike or interburst intervals, and extracting parameters of the underlying chaotic
dynamics for applications of chaos control. Although the issues and the results presented in this work are
relevant for biological systems in general, the current study has focussed on epileptic systems in particular. 
        
Contrary to regular dynamical systems, biological time series are short, noisy, and nonstationary. We find that
amongst all UPO detection methods, the topolological recurrence method \cite{Cvitanovic1988} is most 
reliable for analysis of biological time series for determinism.  It detects, with reasonable success,
statistically significant UPOs in different known dynamical systems with different geometry and in presence of
moderate noise. It is also observed that as long as the added noise does not significantly alter the shape of the 
attractor in the return map, the method is found to be robust and applicable for datasizes exceeding $1000$. 
However, the method was found to be not accurate in determining the fixed point of the detected UPO trajectory. 
This in turn affects the measurement of slopes of the stable and unstable manifolds. At the same time, a dynamical 
transformation method \cite{Soetal1997} is more accurate in determining the location of fixed point in the return map. 
We propose a new hybrid UPO detection method that combines topolological recurrence method for UPO trajectory, 
dynamical transformation for the fixed points, and matrix fit algorithm for slope of the manifolds and this gives 
statsitically significant UPO detection and much improved UPO parameters. For researchers analyzing biological 
time series, this can be a robust mothod for establishing determinism. This also establishes UPO detection as a 
viable tool for establishing determinism in moderate sized noisy biological time series, provided appropriate 
surrogate analysis is adopted for determining the statistical significance of the detected UPOs. 
Although the success of the method can be generalized to systems of different attractor shape and geometry, 
the statistical significance of UPOs depend on the geometry of the UPO trajectory and the frequency of visiting the trajectory.

The method does not, however, provide very accurate UPO parameters for chaos control strategies. Our results,
obtained from known dynamical systems with noise,
suggest that even minor inaccuarcy in the UPO parameters makes the chaos control drift quickly, needing frequent 
external stimuli to keep the system near the targeted periodic orbit. The error is likely to become more serious if the system
is nonstationarity. If the UPOs drift becuase of nonstationarity, the UPO trajectories are to be repeatedly assessed
from short windows. The existing methods are not reliable in determining UPO parameters from short data windows 
if noise is present.     
New adaptive methods have been proposed to overcome this. The repeat chaos control experiements conducted by 
Slutzky {\it et al} \cite{Slutzkyetal2003,Slutzkyetal2002,Slutzkyetal2001} reported to have achieved successful control by adopting stringent adaptive 
tracking and detection techniques. However, our analysis demands a
more careful examination of the procedure and interpretation of the reported control results.  Using a computer model of
epileptic brain slices, the chaos control experiment is repeated using the same adaptive tracking method and produced
comparable results. Since the epileptic bursts timings in the computer model is stochastic in nature, the comparable success 
casts doubt on the interpretation of the results of the experiments on real system. Since the original system also needed frequent 
stimulus, the application strategy is similar to demand pacing, although chaos control strategy that exploits underlying
deterministic dynamics ought to be significantly different from demand pacing.    

The results and analysis presented in this work may not be sufficient to conclude that determinism is positively absent in
epileptic EEG. It simply means that for a sufficiently high dimensional and noisy biological system such as the epileptic
brain slice, the underlying dynamics, even if possibly deterministic, is difficult to unfold with the currently available nonlinear
dynamics tools. The underlying biological complexity and stochastic fluctuations make the task difficult. Nevertheless, non 
linear time series analysis has been seen useful in analysis of such systems. Finding more useful complexity 
measures as well as developing more stringent dynamical tools to unambiguosly describe the dynamics of such biological system continue
to remain a challenging task.

\section{Acknowledgment}
N M  acknowledges a Senior Research Fellowship from Council of
Scientific \& Industrial Research (CSIR), New Delhi. H P S and B B acknowledge
R\&D grant from University of Delhi.

\bibliographystyle{aip}

\bibliography{aip}

\end{document}